\documentclass[12pt]{iopart}

\newcommand{\pabl}[2]{\frac{\partial #1}{\partial #2}}
\newcommand{\ppabl}[3]{\frac{\partial^2#1}{\partial#2\partial#3}}
\newcommand{\PPabl}[2]{\frac{\partial^2#1}{\partial #2^2}}
\newcommand{\beEq}{\stackrel{\textstyle !}{=}}
\newcommand{\sign}{{\rm sign}\,}
\newcommand{\Hess}{{\rm Hess}\,}
\newcommand{\cbrt}[1]{\sqrt[3]{#1}}
\newcommand{\po}{_{\rm po}}
\renewcommand{\Re}{{\rm Re}\,}
\renewcommand{\Im}{{\rm Im}\,}

\begin{document}

\title{Significance of ghost orbit bifurcations in semiclassical spectra}
\author{T Bartsch\dag, J Main\dag\ and G Wunner\ddag}
\address{\dag\ Institut f\"ur Theoretische Physik I,
           Ruhr-Universit\"at Bochum, D-44780 Bochum, Germany}
\address{\ddag\ Institut f\"ur Theoretische Physik und Synergetik,
           Universit\"at Stuttgart, D-70550 Stuttgart, Germany}

\begin{abstract}
Gutzwiller's trace formula for the semiclassical density of states in
a chaotic system diverges near bifurcations of periodic orbits, where
it must be replaced with uniform approximations.  It is well known
that, when applying these approximations, complex predecessors of
orbits created in the bifurcation (``ghost orbits'') can produce
pronounced signatures in the semiclassical spectra in the vicinity of
the bifurcation. It is the purpose of this paper to demonstrate that
these ghost orbits themselves can undergo bifurcations, resulting in
complex, nongeneric bifurcation scenarios. We do so by studying an
example taken from the Diamagnetic Kepler Problem, viz.\ the period
quadrupling of the balloon orbit.  By application of normal form
theory we construct an analytic description of the complete
bifurcation scenario, which is then used to calculate the pertinent
uniform approximation.  The ghost orbit bifurcation turns out to
produce signatures in the semiclassical spectrum in much the same way
as a bifurcation of real orbits would.
\end{abstract}

\pacs{05.45.+b, 03.65.Sq, 32.60.+i}

\section{Introduction}
Since its discovery in the early 1970s, Gutzwiller's trace formula
\cite{Gut67,Gut90} has become a widely used tool for the interpretation of
quantum mechanical spectra of systems whose classical counterpart
exhibits chaotic behaviour. It represents the density of states of the
quantum system as a sum over a smooth part and fluctuations from all 
periodic orbits of the classical system, where the contribution of a 
single periodic orbit reads
\begin{equation}\label{Gutzw}
  {\cal A}\po=\frac{T\po \rme^{\rmi(S\po/\hbar-\frac{\pi}{2}\mu\po)}}
                   {\sqrt{|\det(M\po-I)|}} \;,
\end{equation}
with $T\po, S\po, M\po, \mu\po$ denoting the orbital period, action,
monodromy matrix, and Maslov index, respectively.
This formula assumes that all periodic orbits can be regarded as
isolated, which is the case, in particular, for completely hyperbolic
systems. In the generic case of mixed regular-chaotic dynamics,
however, the formula fails whenever bifurcations of periodic orbits
occur, because close to a bifurcation periodic orbits approach one
another arbitrarily closely. The failure of the formula manifests
itself in divergences of the periodic orbit contributions
(\ref{Gutzw}).

The generic cases of period-$m$-tupling bifurcations were studied by
Ozorio de Almeida and Hannay \cite{Alm87,Alm88}, who derived uniform
semiclassical approximations by taking into account all orbits
involved in a bifurcation collectively.  Their solutions were refined
by Sieber and Schomerus \cite{Sie96,Sch97a,Sie98}, who derived uniform
approximations for all types of bifurcations of codimension one in
generic Hamiltonian systems with two degrees of freedom. Their
formulas smooth the divergence in Gutzwiller's trace formula, and, in
contrast to the approximations in \cite{Alm87,Alm88}, asymptotically
approach the result of the trace formula (\ref{Gutzw}) for isolated
periodic orbit contributions as the distance from the bifurcation
increases.
As a characteristic feature, uniform approximations require the
inclusion of complex ``ghost orbits''. At the bifurcation points, new
periodic orbits are born. However, before they come into being, the
orbits possess predecessors -- ghost orbits -- in the complexified
phase space.  As was shown by Ku\'s et al.\ \cite{Kus93}, some of
these ghost orbits, which in the limit $\hbar\to 0$ yield
exponentially small contributions, have to be included in Gutzwiller's
trace formula (\ref{Gutzw}). As a result, in constructing a uniform
approximation complete information about the bifurcation scenario
including the ghost orbits is required.

A closer inspection of the bifurcation scenarios encountered in
practical applications of uniform approximations reveals that
bifurcations of codimension two, although they cannot generically be
observed if only one control parameter is varied, can nevertheless
have an effect on semiclassical spectra, because in their
neighbourhood two bifurcations of codimension one come close to each
other, and therefore have to be treated collectively. Examples of that
situation have been studied by Main and Wunner \cite{Mai97,Mai98} as
well as Schomerus and Haake \cite{Sch97b,Sch97c}.

The bifurcation scenarios described in the literature so far involve
bifurcations of real orbits only.  However, one should expect
bifurcations of ghost orbits also to be possible and of particular
importance for complicated bifurcation scenarios with codimension
greater than one. It is the purpose of this paper to demonstrate that
ghost orbit bifurcations do indeed occur and have a pronounced effect
on semiclassical spectra.  To this end, we present an example taken
from the Diamagnetic Kepler Problem. It turns out that even the
analysis of the period-quadrupling bifurcation of one of the shortest
periodic orbits in that system requires the inclusion of a ghost
bifurcation.

The appearance of ghost orbit bifurcations represents an additional
challenge for the construction of uniform approximations. It will turn
out that normal form theory allows treating both ghost bifurcations
and bifurcations of real orbits on an equal footing. Consequently,
ghost bifurcations contribute to uniform approximations in much the
same way as real bifurcations do. From these observations we conclude
that the occurence of ghost bifurcations in systems with mixed
regular-chaotic dynamics is not a very exotic but rather quite a
common phenomenon.

The organization of the paper is as follows: In \sref{BifSec} we
describe the bifurcation scenario of the example chosen in detail.
\Sref{UnifSec} provides the general form of uniform approximations.
\Sref{NFSec} presents the normal form describing the bifurcation
scenario in point and the discussion of how the ghost bifurcation can
be included in the normal form.  In section \ref{ParmSec} we determine
the normal form parameters to quantitatively describe the
bifurcations, and in \sref{EvalSec} the uniform approximation is
evaluated.

\section{The bifurcation scenario}
\label{BifSec}
As an example, we study the hydrogen atom in a magnetic field, which
has been described in detail, e.g., in Refs.\
\cite{Fri89,Has89,Wat93}.  We assume the nucleus fixed and regard the
electron as a structureless point charge.  If the magnetic field is
directed along the $z$-axis, the nonrelativistic Hamiltonian
describing the electron motion reads [in atomic units, with
$\gamma=B/(2.35\times 10^5\,{\rm T})$ the magnetic field strength]
\begin{equation}
  H = \frac{1}{2}\bi{p}^2 +\frac{1}{2}\gamma L_z 
     +\frac{1}{8}\gamma^2\left(x^2+y^2\right)-\frac{1}{r} = E \;.
\end{equation}
Here, $r$ is the distance from the nucleus, and $L_z$ denotes the
angular momentum along the field axis, which is conserved because of
the rotational symmetry around that axis.  In the following we
restrict ourselves to the case where $L_z = 0$.

To further simplify the Hamiltonian, we exploit its scaling property
with respect to the magnetic field strength $\gamma$. 
In scaled coordinates and momenta
\begin{equation}\label{prScal}
  \tilde{\bi{r}} = \gamma^{2/3}\bi{r},\qquad
  \tilde{\bi{p}} = \gamma^{-1/3}\bi{p} 
\end{equation}
the Hamiltonian assumes the form
\begin{equation}\label{hamScal}
  \tilde{H} = \gamma^{-2/3} H 
    = \frac{1}{2}\tilde{\bi{p}}^2 
     + \frac{1}{8} \left(\tilde{x}^2+\tilde{y}^2\right)
     - \frac{1}{\tilde{r}} = \tilde{E} \;.
\end{equation}
Thus, the classical dynamics does not depend on the energy $E$ and field
strength $\gamma$ separately, but only on the scaled energy $\tilde{E}
= \gamma^{-2/3}E$. From the scaling prescriptions (\ref{prScal}) and
(\ref{hamScal}) we derive the scaling laws for classical actions and
times as
\begin{equation}
  \tilde{S} = \gamma^{1/3}S\;,\qquad \tilde{T}=\gamma T\;.
\end{equation}

Due to the Coulomb potential, the Hamiltonian (\ref{hamScal}) is singular 
at $\tilde r=0$.
The equations of motion can be regularized by introducing semiparabolical
coordinates 
\begin{equation}\label{spKoord}
  \mu^2 = \tilde{r}+\tilde{z}\;,\qquad
  \nu^2 = \tilde{r}-\tilde{z} \; ,
\end{equation}
and a new time parameter $\tau$ defined by $dt = 2 r\, d\tau$.
Finally, the regularized Hamiltonian is obtained as \cite{Fri89,Has89,Wat93}
\begin{equation}
  {\cal H}= \frac{1}{2}\left(p_\mu^2+p_\nu^2\right)
           - \tilde{E}\left(\mu^2+\nu^2\right)
           + \frac{1}{8}\mu^2\nu^2\left(\mu^2+\nu^2\right) \equiv 2\; ,
\end{equation}
and Hamilton's equations of motion read
[with primes denoting derivatives $\rmd/\rmd\tau$],
\begin{eqnarray} \label{HamEqs}\eqalign{
  \mu' = p_\mu\;, \qquad& p_\mu' = 2\tilde{E}\mu 
                        -\frac{1}{4}\mu\nu^2(2\mu^2+\nu^2)\;, \\
  \nu' = p_\nu\;, & p_\nu' = 2\tilde{E}\nu
                        -\frac{1}{4}\mu^2\nu(\mu^2+2\nu^2)\;.
}\end{eqnarray}
These equations are free of singularities and can easily be integrated
numerically. When using them, we must keep in mind that 
the definition (\ref{spKoord}) determines the semiparabolical
coordinates $\mu, \nu$ up to a choice of sign only, giving a
many-to-one coordinate system. Thus, if we integrate the equations of
motion (\ref{HamEqs}) until the trajectory closes in
$(\mu,\nu)$-coordinates this may correspond to more than one period
in the original configuration space. Furthermore, we have to identify
orbits which can be transformed into one another by reflections at the
coordinate axes.

We now complexify our phase space by allowing coordinates and momenta
to assume complex values. This extension allows us to look for ghost
predecessors of real orbits born in a bifurcation \cite{Mai97}.

At any given scaled energy $\tilde{E}$, there is a periodic orbit
along the magnetic field axis. For sufficiently low negative
$\tilde{E}$ it is stable, while, as $\tilde{E}\nearrow 0$, it loses
and regains stability infinitely often \cite{Win87,Fri89}.  Stability
is lost, for the first time, at $\tilde{E} =-0.391$.  At this scaled
energy, the so-called balloon orbit \cite{Mao92} is born as a new,
stable periodic orbit.  As the scaled energy increases further, the
orbit exhibits all kinds of period-$m$-tupling bifurcations before
finally turning unstable at $\tilde{E}=-0.291$.

\begin{figure}
\vspace{20.0cm}
\includegraphics{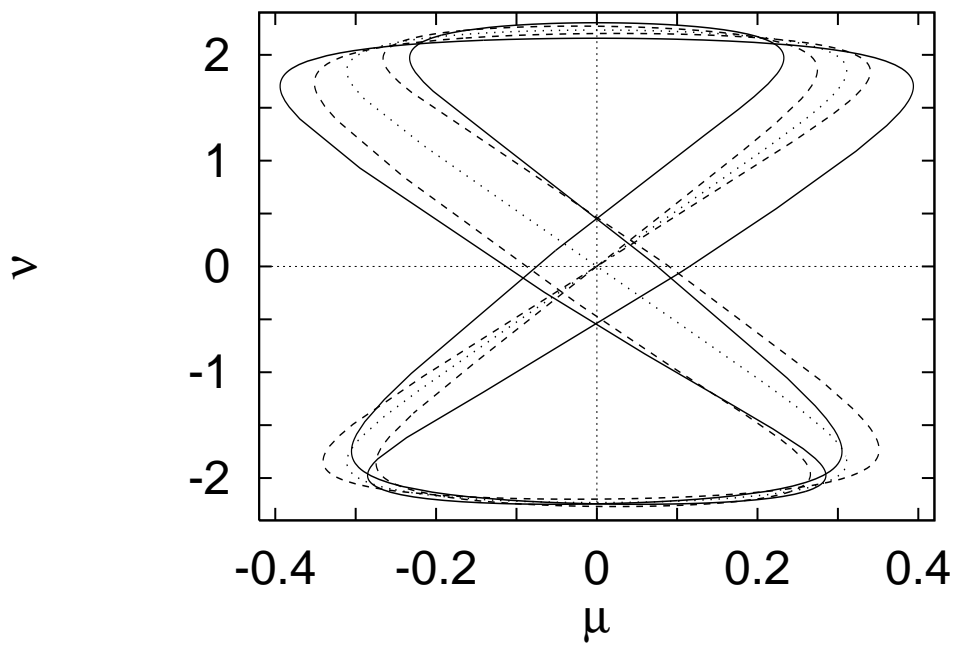}
\caption{\label{reSat}
Real orbits at scaled energy $\tilde{E}=-0.340>\tilde E_c$ drawn in
scaled semiparabolical coordinates 
$(\mu=\gamma^{1/3}(r+z)^{1/2},\nu=\gamma^{1/3}(r-z)^{1/2})$.
Solid and dashed lines: Stable and unstable satellites created at the
period-quadrupling bifurcation of the balloon orbit.
Dotted line: Balloon orbit.}
\end{figure}

\begin{figure}
\vspace{20.0cm}
\includegraphics{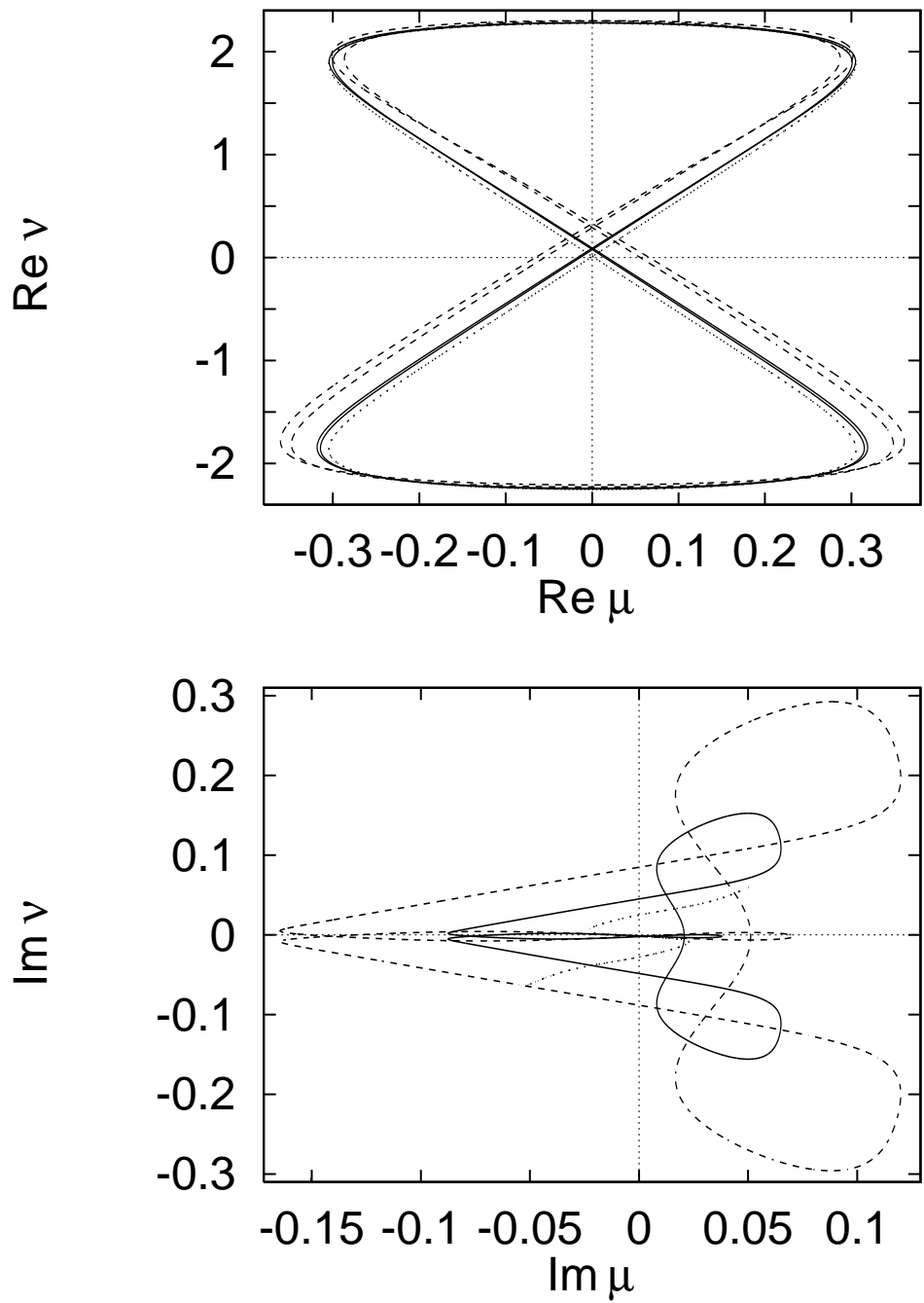}
\caption{\label{ghSat1}
Real and imaginary parts of complex ghost orbits at scaled energy 
$\tilde E=-0.343$. Solid and dotted lines: Stable and unstable ghost 
satellites created at the period-quadrupling bifurcation of the balloon
orbit at $\tilde E_c=-0.342025$.
Dashed line: Additional ghost orbit created at the ghost bifurcation
at $\tilde E_c'=-0.343605$.}
\end{figure}

Here, we consider the period-quadrupling bifurcation of the balloon
orbit at $\tilde{E}_c=-0.342025$.  For $\tilde{E}>\tilde{E}_c$, two
real satellite orbits of quadruple period exist.  As
$\tilde{E}\searrow\tilde{E}_c$, they collide with the balloon orbit
and form an island-chain bifurcation as described in
\cite{Alm87,Alm88}.  The real orbits are shown in figure \ref{reSat}
at scaled energy $\tilde E=-0.340$. The solid and dashed curves
represent the stable and unstable satellite orbits, respectively. For
comparison, the balloon orbit is shown as a dotted curve. Below
$\tilde{E}_c$, no real satellite orbits exist.  Instead, there are a
stable and an unstable complex ghost satellite. The real and imaginary
parts of the stable and unstable ghost orbits at scaled energy $\tilde
E=-0.343$ are drawn as solid and dotted curves in figure \ref{ghSat1}.
Note that the imaginary parts are small compared to the real parts.
As predicted by normal form theory (see \sref{generic_bif}), both
satellites coincide with their complex conjugates, whence the total
number of orbits is conserved in the bifurcation.

\begin{figure}
\vspace{20.0cm}
\includegraphics{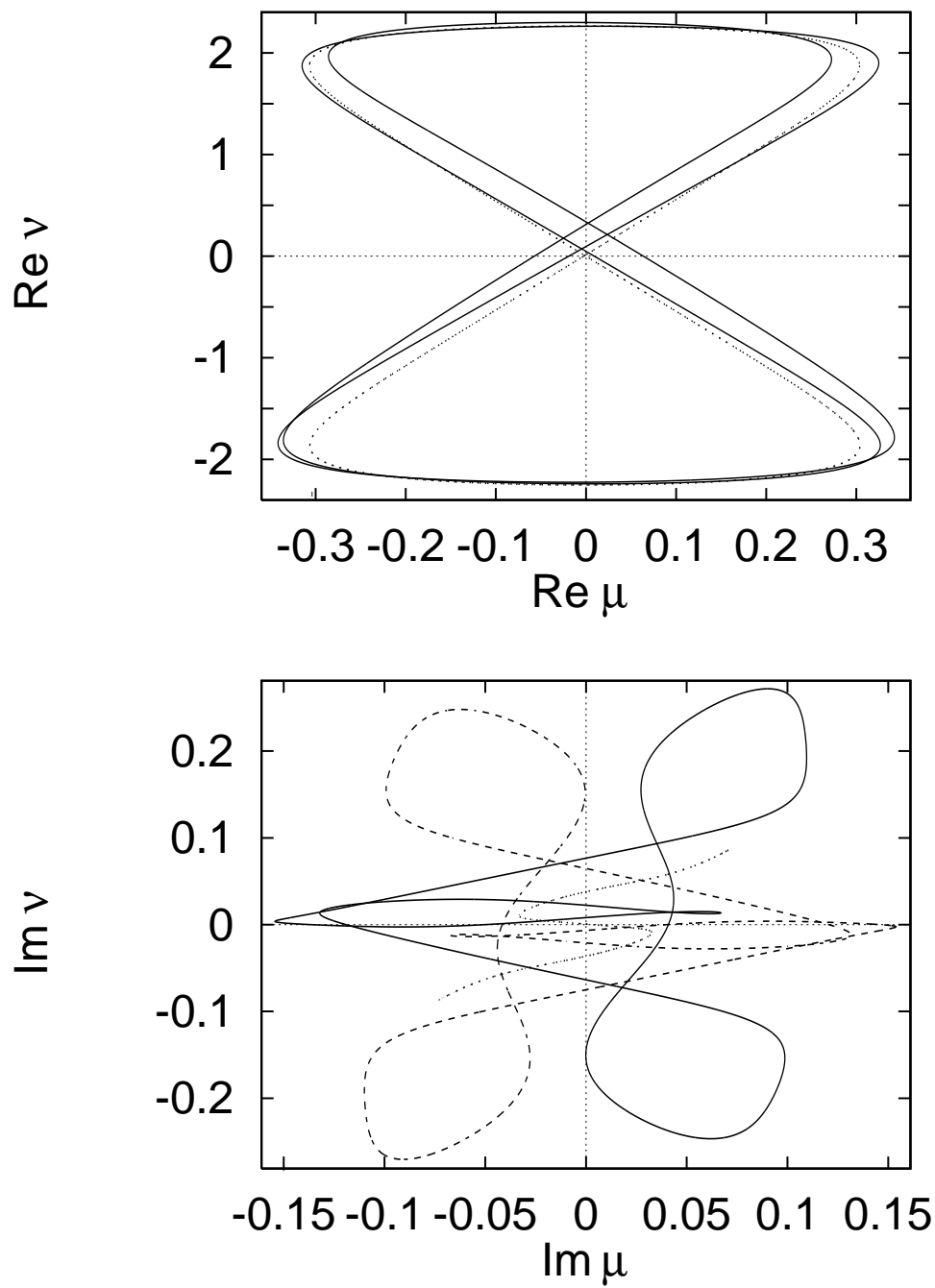}
\caption{\label{ghSat2}
Ghost orbits at scaled energy $\tilde{E}=-0.344<\tilde{E}_c'$. 
Solid and dashed lines: Asymmetric ghosts (real parts coincide) created
at the ghost bifurcation at $\tilde E_c'=-0.343605$.
Dotted line: Unstable ghost satellite created at the period-quadrupling 
bifurcation of the balloon orbit at $\tilde E_c=-0.342025$.}
\end{figure}

The orbits described so far form a generic type of period-quadrupling
bifurcation as investigated by Sieber and Schomerus \cite{Sie98}.
However, the classical periodic orbit search in the complexified phase
space reveals the existence of an additional ghost orbit at scaled
energies around $\tilde E_c$.  The shape of this orbit is shown as the
dashed curve in figure \ref{ghSat1}.  It is similar to the stable
ghost satellite originating from the period-quadrupling of the balloon
orbit.  When following this ghost orbit to lower energies we find
another bifurcation at $\tilde E_c'=-0.343605$, i.e., slightly below
the bifurcation point $\tilde E_c=-0.342025$ of the period-quadrupling
of the balloon orbit.  At energy $\tilde E=\tilde E_c'$ the additional
ghost orbit (dashed curve in figure \ref{ghSat1}) collides with the
stable ghost satellite of the period-quadrupling bifurcation (solid
curve in figure \ref{ghSat1}), and these two orbits turn into a pair
of complex conjugate ghost orbits.  Their shapes are presented at
scaled energy $\tilde E=-0.344$ by the solid and dashed curves in
figure \ref{ghSat2}.  The imaginary parts clearly exhibit the loss of
conjugation symmetry described, if the aforementioned symmetry of the
semiparabolical coordinate system is taken into account.  The dotted
curve in figure \ref{ghSat2} is the unstable ghost satellite which
already exists at higher energy $\tilde E>\tilde E_c'$ (see the dotted
curve in figure \ref{ghSat1}).

It is important to note that the second bifurcation at $\tilde
E=\tilde E_c'$ involves ghost orbits only.  This kind of bifurcation
has not been described in the literature so far; in particular,
Meyer's classification of codimension-one bifurcations \cite{Mey70}
contains bifurcations of real orbits only and does not cover ghost
bifurcations.  The existence of the ghost orbit bifurcation implies
that the results of Ref.\ \cite{Sie98} for the uniform semiclassical
approximation for the generic period-quadrupling bifurcation cannot be
applied to the more complicated bifurcation scenario considered here.
As in cases described before by Main and Wunner \cite{Mai97,Mai98} and
Schomerus and Haake \cite{Sch97b,Sch97c}, the closeness of the two
bifurcations requires the construction of a uniform approximation taking
into account all orbits involved in the successive bifurcations
collectively.  Thus, the ghost bifurcation at $\tilde{E}_c'$ turns out
to contribute to the semiclassical approximation in the same way as a
real bifurcation would, as long as we do not go to the extreme
semiclassical domain where the bifurcations can be regarded as
isolated.  We will demonstrate in \sref{NFSec} that the techniques of
normal form theory can be extended as to include the description of
ghost bifurcations.

\begin{figure}
\vspace{20.0cm}
\includegraphics{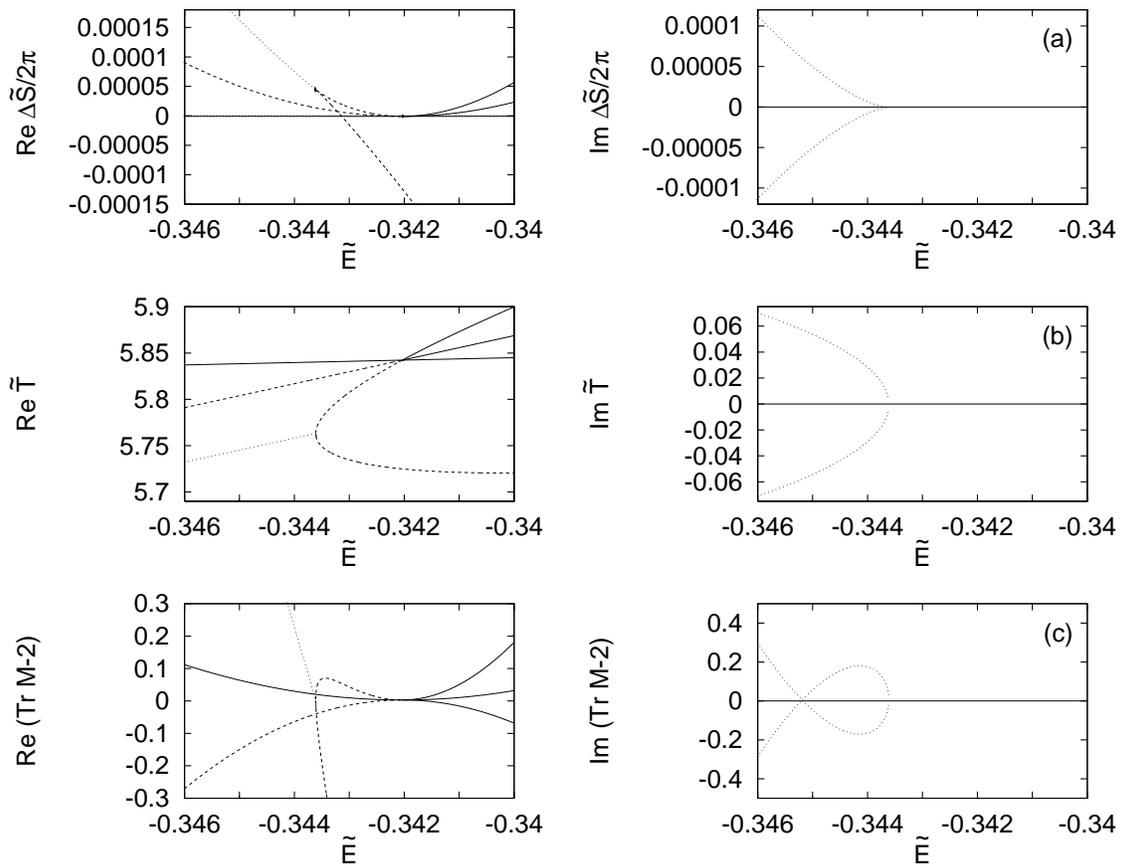}
\caption{\label{orbDat}Actions, periods and traces of the orbits
involved in the bifurcations as functions of the scaled energy
$\tilde E=E\gamma^{-2/3}$.
Solid lines: real orbits, dashed lines: ghost orbits symmetric with 
respect to complex conjugation, dotted lines: asymmetric ghost orbits.}
\end{figure}

The construction of the uniform approximation requires the knowledge
of the periodic orbit parameters of all orbits participating in the
bifurcation scenario.  The numerically calculated parameters are shown
in \fref{orbDat} as functions of the scaled energy $\tilde E$.  Part
(a) of this figure displays the actions of the periodic orbits, where
the action of four repetitions of the balloon orbit was chosen as a
reference level ($\Delta S =0$).  This kind of presentation exhibits
the sequence of bifurcations more clearly than a plot of the actual
action integrals. Around $\tilde{E}_c$, we recognize two almost
parabolic curves, indicating the actions of the stable (upper curve)
and unstable (lower curve) satellites, respectively. At $\tilde{E}_c$,
the curves change from solid to dashed as the satellites become
complex. Below $\tilde{E}_c$, the unstable ghost satellite does not
undergo any further bifurcations in the energy range shown, whereas
the stable satellite collides, at $\tilde{E}_c'$, with the additional
ghost orbit, which can clearly be seen not to be involved in the
bifurcation at $\tilde{E}_c$. Below $\tilde{E}_c'$, these two orbits
are complex conjugates of each other. Thus, the real parts of their
actions coincide, whereas the imaginary parts are different from zero
and have opposite signs.

Analogously, \fref{orbDat}b displays the orbital periods.  Here, no
differences were taken, so that the fourth repetition of the balloon
orbit, which is always real, shows up as a nearly horizontal solid
line at $\tilde T\approx 5.84$.  The other orbits can be identified
with the help of the bifurcations they undergo, in the same way as
discussed above.  Finally, \fref{orbDat}c presents the traces of
monodromy matrices minus two.  These quantities agree with $\det(M-I)$
for systems with two degrees of freedom.  At $\tilde{E}_c$ and
$\tilde{E}_c'$, they can be seen to vanish for the orbits involved in
the bifurcations, causing the divergences of the periodic orbit
amplitudes \eref{Gutzw} at the bifurcation points.

The generic period-quadrupling bifurcation at $\tilde{E}_c$ can be
described with the help of lowest order normal form theory presented
in \sref{generic_bif}.  As the ghost bifurcation is approached, this
description fails.  However, the influence of the additional ghost
orbit can be taken into account by including higher order terms in the
normal form as discussed in \sref{ungeneric_bif}.

\section{The general form of the uniform approximation}
\label{UnifSec}

Before we return to classical normal form theory in \sref{NFSec},  
we introduce, in this section,
the basic formulas for the quantum density
of states necessary for the construction of the uniform semiclassical
approximation.
The density of states of a quantum system with the Hamiltonian $\hat H$
can be expressed with the help of the Green's function $G(E)=(E-\hat
H)^{-1}$ as
\begin{equation}
  d(E) = -\frac{1}{\pi}\Im \Tr \, G(E) \; ,
\end{equation}
where the trace of the Green's function can be evaluated in the
coordinate re\-pre\-sen\-ta\-tion,
\begin{equation}
   \Tr \, G(E)
 = \int d\bi{x'}d\bi{x}\delta(\bi{x'}-\bi{x}) G(\bi{x'}\bi{x},E) \; .
\end{equation}
The basic steps in the formulation of {\em periodic orbit theory}
\cite{Gut67,Gut90} are to replace the Green's function
$G(\bi{x'}\bi{x},E)$ with its semiclassical Van Vleck-Gutzwiller
approximation and to carry out the integrals in stationary-phase
approximation.  For systems with two degrees of freedom the
semiclassical approximation to the Green's function reads
\begin{equation}
\fl G(\bi{x'}\bi{x}, E) = \frac{1}{\rmi\hbar\sqrt{2\pi\rmi\hbar}}
     \sum_{\rm class.~traj.}\sqrt{D}
     \exp\left\{\frac{\rmi}{\hbar}S(\bi{x'}\bi{x}, E)
                 -\rmi\frac{\pi}{2}\nu\right\} \; .
\end{equation}
Here, the sum extends over all classical trajectories running from
$\bi{x}$ to $\bi{x'}$ at energy $E$, $S$ is the action of a
trajectory, $\nu$ its Maslov index, and $D$ is defined by the second
derivatives of the action,
\begin{equation}
  D = \det \left(
     \begin{array}{cc}
       \ppabl{S}{\bi{x'}}{\bi{x}} & \ppabl{S}{\bi{x'}}{E} \\[.5ex]
       \ppabl{S}{E}{\bi{x}} & \PPabl{S}{E}
     \end{array}\right) \; .
\end{equation}
The contribution of a single orbit to the density of states can be
evaluated by introducing coordinates parallel and perpendicular to the
orbit.  The integration along the orbit can then be performed in a
straightforward fashion.  Finally, Gutzwiller's trace formula
(\ref{Gutzw}) for isolated periodic orbits is obtained by integrating
over the coordinates perpendicular to the trajectory using the
stationary-phase approximation.  It is this last step which fails
close to a bifurcation, where periodic orbits are not isolated.

To find an expression for the density of states which is valid close
to bifurcations, it is convenient to go over to a coordinate-momentum
representation of the Green's function.
Close to a period-$m$-tupling bifurcation of a real orbit
the uniform approximation takes the form \cite{Sie96}
\begin{equation}
\fl d(E) = \frac{1}{2\pi^2m\hbar^2}\, \Re
           \int dy\,dp_y'\, \pabl{\hat{S}}{E}
            \sqrt{\left|\ppabl{\hat{S}}{y}{p_y'}\right|}
        \exp\left\{\frac{\rmi}{\hbar}\left(\hat{S}+yp_y'\right)
                    -i\frac{\pi}{2}\hat{\nu}\right\} \; ,
\label{d_uni}
\end{equation}
with $y$, $p_y'$, and $\hat S$ defined as follows.
Let $(y,p_y)$ be the canonical variables in the Poincar\'e surface 
of section perpendicular to the bifurcating orbit, and $(y',p_y')$ 
the corresponding variables after the period-$m$ cycle of the orbit.
The function
\begin{equation}
  \hat{S}(p_y' y,E) = S(y' y,E) -y'p_y'
\end{equation}
is the Legendre transform of the action integral with respect to the
final coordinate or, in other words, the generating function of the
Poincar\'e map for $m$ periods of the bifurcating orbit in a
coordinate-momentum representation. 

The function in the exponent in (\ref{d_uni}),
\begin{equation}
  f(y,p_y',E) = \hat{S}(y,p_y',E) +y p_y' \;,
\end{equation}
is stationary at the fixed points of the $m$-traversal Poincar\'e map,
that is, fixed points of $f$ correspond to classical periodic
orbits. In the spirit of catastrophe theory, we now relate $f$ to a
given ansatz function $\Phi$, which has the same distribution of
stationary points, by a smooth invertible change of coordinates $\psi$
as
\begin{equation}
  f(y,p_y',E) = S_0(E) +\Phi(\psi(y,p_y';E);E) \;.
\end{equation}
Here, $\psi$ is assumed to keep the origin fixed, $\psi(0,0;E)=(0,0)$, 
and $S_0(E)$ is the action of the central bifurcating orbit.
Inserting this ansatz, we obtain
\begin{eqnarray}\label{unifApprox}\eqalign{
\fl d(E) = \frac{1}{2\pi^2m\hbar^2}\, \Re 
         \exp\left\{\frac{\rmi}{\hbar}S_0(E)
                    -\rmi\frac{\pi}{2}\hat{\nu}\right\}\\
         \times \int dY\,dP_Y' \pabl{\hat{S}}{E}
         \sqrt{\left|\ppabl{\hat{S}}{y}{p_y'}\right|} 
         \sqrt{\frac{|\Hess\Phi|}{|\Hess f|}}
         \exp\left\{\frac{\rmi}{\hbar}\Phi(Y,P_Y')\right\}\;,
}\end{eqnarray}
where $\Hess$ denotes the Hessian matrix and the coefficient
\begin{equation}
  X := \pabl{\hat{S}}{E}
         \sqrt{\left|\ppabl{\hat{S}}{y}{p_y'}\right|} 
         \sqrt{\frac{|\Hess\Phi|}{|\Hess f|}}
\end{equation}
is unknown. At a stationary point of $\Phi$, it can be shown to assume
the value
\begin{equation}
  X \stackrel{sp}{=} \frac{\{m\}T}{\sqrt{|\Tr M-2|}}\sqrt{|\Hess\Phi|}\;,
\end{equation}
where $T$ and $M$ denote the period and the monodromy matrix of the
corresponding classical orbit, and the notation $\{m\}$ is meant to
indicate that this factor does not occur at the satellite orbits.

\section{Normal-form description of the bifurcation} 
\label{NFSec}
To evaluate the uniform approximation \eref{unifApprox}, we need to find 
a suitable ansatz function $\Phi$.
Normal form theory as developed by Birkhoff \cite{Bir27} and Gustavson 
\cite{Gus66} provides us with a systematic way to construct such ansatz 
functions. 
We will first investigate the lowest nontrivial order of the normal form 
expansion which describes generic bifurcations. 
Then, we will show that higher-order terms in the expansion can account 
for the more complicated bifurcation scenario studied here.

\subsection{The generic period-quadrupling bifurcation}
\label{generic_bif}
The simplest normal form
describing the generic period-quadrupling bifurcations reads \cite{Alm87}
\begin{equation}\label{NF2}
  \Phi = \varepsilon I + a I^2 + bI^2\cos(4\varphi) \;.
\end{equation}
This normal form is expressed in terms of canonical (action-angle)
polar coordinates
$(I,\varphi)$, which are connected to Cartesian coordinates $(p,q)$ by
\begin{equation}\label{polKoord}
  p = \sqrt{2I}\cos\varphi\;,\qquad q=\sqrt{2I}\sin\varphi\;.
\end{equation}

To establish the connection with classical periodic orbits, we need to
determine the stationary points of the normal form and then evaluate
its stationary values. The stationary points are given by the
equations
\begin{eqnarray}\eqalign{
  0 &\beEq \pabl{\Phi}{\varphi} = -4I^2b\sin(4\varphi) \;,\\
  0 &\beEq \pabl{\Phi}{I} = \varepsilon +2aI+2bI\cos(4\varphi) \;.
}\end{eqnarray}
The first of these equations yields
\begin{eqnarray}\eqalign{
  \sin(4\varphi) = 0 \;, \\
  \cos(4\varphi) = \sigma = \pm 1\;,
}
\label{stat_phi}
\end{eqnarray}
so that the second equation reads
\begin{equation*}
  \varepsilon + 2(a+\sigma b)I = 0 \; ,
\end{equation*}
with its solution
\begin{equation}
  I_{\sigma} = -\frac{\varepsilon}{2(a+\sigma b)} \; .
\end{equation}
In addition, there is the central periodic orbit at $I=0$, which does
not show up as a stationary point because the polar 
coordinate system (\ref{polKoord}) is singular at the origin.

To interprete these results, we observe that according to its
definition (\ref{polKoord}) the coordinate $I$ is positive for real
orbits and that the action difference $\Phi(I,\varphi)$ is real for
real $I, \varphi$. Therefore, negative real solutions $I$ correspond
to ghost orbits which are symmetric with respect to complex
conjugation and thus have real actions, whereas a complex $I$ indicates
an asymmetric ghost orbit.

Now, if $|a|>|b|$, the two solutions $I_{\pm}$ have the same sign,
which changes at $\varepsilon=0$. So, the two satellite orbits change
from two real orbits to two ghosts at the bifurcation point
$\varepsilon=0$, forming an island-chain-bifurcation.
If $|a|<|b|$, $I_+$ and $I_-$ have different signs, so that on either
side of the bifurcation there is one real and one ghost satellite. At
$\varepsilon=0$, the satellites collide with the central orbit,
forming a touch-and-go-bifurcation.

\subsection{Generalization to nongeneric bifurcations}
\label{ungeneric_bif}
To describe a sequence of two bifurcations, we need to include
higher-order terms in the normal form.  Here, we adopt the normal
form
\begin{equation}\label{NF}
  \Phi = \varepsilon I + a I^2 + bI^2\cos(4\varphi)
        + cI^3(1+\cos(4\varphi))
\end{equation}
given by Schomerus \cite{Sch98} as a variant of the normal form used
by Sadovski\'\i\ and Delos \cite{Sad96} to describe a sequence of
bifurcations close to a period-quadrupling. It turns out to
qualitatively describe the sequence of bifurcations encountered here
for suitably chosen parameter values.

The stationary points of $\Phi(I,\varphi)$ are given by the equations
\begin{eqnarray}\eqalign{
  0 &\beEq \pabl{\Phi}{\varphi} = -4I^2(b+cI)\sin(4\varphi) \;,\\
  0 &\beEq \pabl{\Phi}{I} = \varepsilon +2aI+2bI\cos(4\varphi)
            +3cI^2(1+\cos(4\varphi)) \;.
}\end{eqnarray}
As in \sref{generic_bif}, the first of these equations yields
(see eq. \ref{stat_phi})
\begin{eqnarray*}
  \sin(4\varphi) = 0 \;, \\
  \cos(4\varphi) = \sigma = \pm 1 \; .
\end{eqnarray*}
Thus, the second equation reads
\begin{equation}
  \varepsilon + 2(a+\sigma b)I + 3cI^2(1+\sigma) = 0\;.
\end{equation}
In solving this equation, we shall assume $c<0$. 

For $\sigma=-1$, the equation is linear. Its only solution reads
\begin{equation}
  I_{-1} = -\frac{\varepsilon}{2(a-b)}\;.
\end{equation}
For $\sigma=+1$, however, we obtain a quadratic equation with two
solutions:
\begin{equation}
  I_\pm = -c^{-1/3}\left(\delta\pm\sqrt{\eta+\delta^2}\right) \;,
\end{equation}
where we have introduced the abbreviations
\begin{equation}\label{etaDeltaDef}
  \eta \equiv -\frac{\varepsilon}{6c^{1/3}}, \quad
  \delta \equiv \frac{a+b}{6c^{2/3}} \; .
\end{equation}
Thus, we obtain three different stationary points corresponding to the
three satellite orbits in addition to the central periodic orbit at
$I=0$.

\begin{figure}
\vspace{20.0cm}
\includegraphics{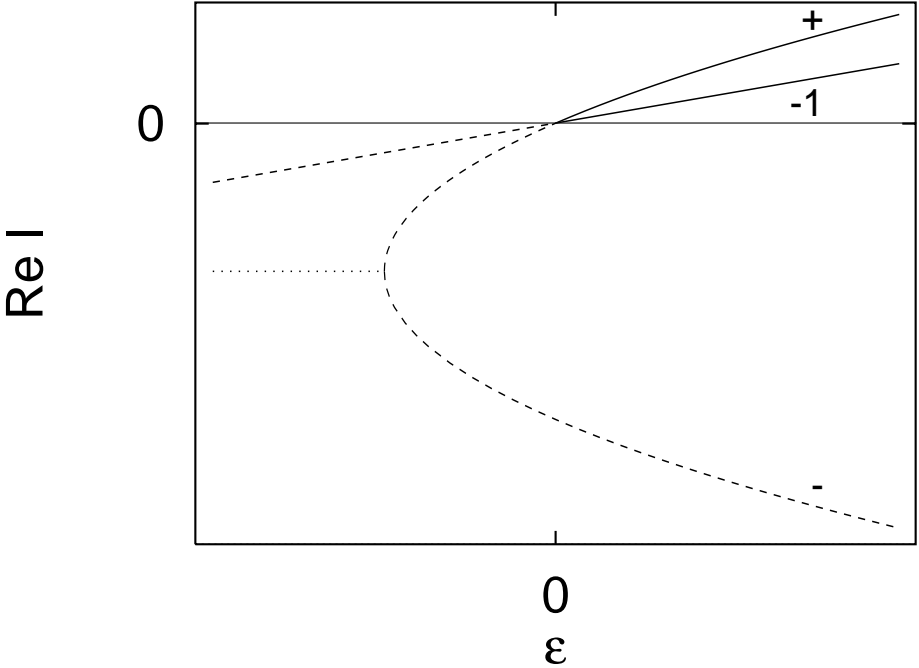}
\caption{\label{IEtaFig}
Sketch of the bifurcation scenario given by the normal form (\ref{NF}) 
for the case $|a|>|b|$ and $c<0$. 
Solid lines: Real orbits. 
Dashed lines: Ghost orbits symmetric with respect to complex conjugation.
Dotted line: A pair of complex conjugate ghosts.}
\end{figure}

The dependence of the solutions $I$ on $\varepsilon$ is shown
schematically in figure \ref{IEtaFig} for the case $|a|>|b|$ and
$c<0$. Comparing the normal form results to the bifurcation scenario
described in section \ref{BifSec}, we recognize the sequence of an
island-chain-bifurcation at $\varepsilon=0$ and a ghost orbit
bifurcation at some negative value of $\varepsilon$. Thus, our normal
form correctly describes the bifurcation scenario under consideration,
and we adopt it as an ansatz function in the uniform approximation
(\ref{unifApprox}). This correspondence allows us to identify
stationary points with classical periodic orbits as follows: The
$(\sigma=-1)$-solution describes the unstable satellite orbits on
either side of $\tilde{E}_c$. For $\tilde{E}<\tilde{E}_c'$, the
$(\sigma=+1)$-solutions correspond to the asymmetric ghost orbits,
whereas for $\tilde{E}>\tilde{E}_c'$, we identify the two solutions 
for $\sigma=+1$ with the stable satellite orbit (marked by the $+$ 
in figure \ref{IEtaFig}) and with the additional ghost ($-$ in figure 
\ref{IEtaFig}).

We now calculate the stationary values of the normal form, which
correspond to action differences. They read
\begin{eqnarray}\label{NFWirk}\eqalign{
  \Phi_\pm = 4\left(\eta+\delta^2\right)
               \left(\delta\pm\sqrt{\eta+\delta^2}\right)+2\eta\delta \; , \\
  \Phi_{-1} = -\frac{\varepsilon^2}{4(a-b)} \; .
}\end{eqnarray}
Finally, we need the Hessian determinants of the normal form at the
stationary points. We calculate them with respect to cartesian
coordinates, because in polar coordinates the Hessian determinant at
the central orbit $I=0$ is undefined, and obtain
\begin{eqnarray}\label{NFHess}\eqalign{
  \Hess_\pm = \left\{\varepsilon+2(a-3b)I_\pm-2cI_\pm^2\right\}
              \left\{\varepsilon+6(a-b)I_\pm+30cI_\pm^2\right\} \;, \\
  \Hess_{-1}= \left\{\varepsilon+4aI_{-1}+4cI_{-1}^2\right\}^2
              -4I_{-1}^2\left\{a-3b-2cI_{-1}\right\}^2 \;, \\
  \Hess_0 = \varepsilon^2 \; .
}\end{eqnarray}

We have now found a normal form which can serve as an analytical
description of the complicated bifurcation scenario.

\section{Determination of normal form parameters}
\label{ParmSec}
In order to use the normal form (\ref{NF}) as an ansatz in the uniform
approximation (\ref{unifApprox}), we now have to determine the
normal form parameters $\varepsilon, a, b,c$ so that the numerically
observed action differences are quantitatively reproduced by the normal
form results (\ref{NFWirk}). Since, according to figure \ref{IEtaFig},
$\varepsilon$ measures the distance from the period quadrupling
bifurcation, we choose
\begin{equation}
  \varepsilon = \tilde{E} - \tilde{E}_c \; ,
\end{equation}
and then solve equations (\ref{NFWirk}) for the parameters $a,b,c$.

To achieve this, we introduce 
\begin{eqnarray}\label{hpmDef}\eqalign{
  h_+ &=\frac{\Phi_++\Phi_-}{8}
        =\delta\left(\eta+\delta^2\right)+\frac{1}{2}\eta\delta \;, \\
  h_- &=\frac{\Phi_+-\Phi_-}{8}
        =(\eta+\delta^2)^{3/2} \;.
}\end{eqnarray}
The second equation gives
\begin{equation}\label{etaVonDelta}
  \eta = h_-^{2/3}-\delta^2 \;,
\end{equation}
so that from the first equation we obtain
\begin{equation}\label{deltaGlg}
  \delta^3-3h_-^{2/3}\delta+2h_+ = 0 \;.
\end{equation}
This is a cubic equation for $\delta$. Its discriminant reads
\begin{equation}
  D = h_+^2-h_-^2 = \frac{1}{16}\Phi_+\Phi_- \;,
\end{equation}
and, using Cardani's formula, we find its solutions
\begin{equation}\label{deltaLsg}
  \delta = \frac{\lambda}{2}
              \cbrt
                   {-\left(\sqrt{\Phi_+}+\sqrt{\Phi_-}\right)^2}
           +\frac{\lambda^\ast}{2}
              \cbrt
                   {-\left(\sqrt{\Phi_+}-\sqrt{\Phi_-}\right)^2} \;,
\end{equation}
where $\lambda\in\left\{1,-\frac{1}{2}\pm
\rmi\frac{\sqrt{3}}{2}\right\}$ is a cube root of unity. If $D>0$,
$\lambda = 1$ yields the only real solution, whereas for $D<0$ all
solutions are real. To proceed, we have to choose one of the three
solutions.

As can be seen from figure \ref{orbDat} using the correspondence
between stationary points and periodic orbits discussed above, we have
$\Phi_+>0$, and there is an $\varepsilon_0<0$ such that $\Phi_->0$ for
$\varepsilon>\varepsilon_0$ and $\Phi_-<0$ for
$\varepsilon<\varepsilon_0$. Consequently, $D>0$ for
$\varepsilon>\varepsilon_0$ and $D<0$ for
$\varepsilon<\varepsilon_0$. To make $\delta$ real, we therefore have
to choose $\lambda = 1$ if $\varepsilon<\varepsilon_0$.

To determine $\lambda$ for $\varepsilon>\varepsilon_0$, we first
observe that the parameters must depend on $\varepsilon$
continuously. Thus, $\lambda$ can only change at energies where
equation (\ref{deltaGlg}) has a double root, viz.\ $D=0$ or
$\varepsilon \in \{0, \varepsilon_0\}$. Therefore, it suffices to
determine $\lambda$ in a neighbourhood of $\varepsilon = 0$.

From the plot of action differences in figure \ref{orbDat} we find
\begin{eqnarray*}
  \Phi_+=\alpha^2\varepsilon^2+\Or\left(\varepsilon^3\right) \; , \\
  \Phi_-=-\Gamma-\beta\varepsilon+\Or\left(\varepsilon^3\right) \; ,
\end{eqnarray*}
with positive constants $\alpha, \beta, \Gamma$. With the help of
equations (\ref{deltaLsg}) and (\ref{etaVonDelta}) we can now expand
$\eta$ in a Taylor series to first order in $\varepsilon$:
\begin{eqnarray}
\fl \eta
   =\left(\frac{1}{4}-(\Re\lambda)^2\right)\Gamma^{2/3} \nonumber \\
     +\left(\left(\frac{1}{4}-(\Re\lambda)^2\right)
                                 \frac{2\beta}{3\Gamma^{2/3}}
        -\Re\lambda\,\Im\lambda\frac{4\alpha\sqrt{\Gamma}}{3\Gamma^{1/3}}
                   \sign\varepsilon\right)\varepsilon
      +\Or\left(\varepsilon^2\right) \; .
\end{eqnarray}
Requiring this result to reproduce the definition
\[
  \eta = -\frac{1}{6c^{1/3}}\varepsilon\;,\qquad-\frac{1}{6c^{1/3}}>0\;,
\]
we find the conditions
\[\Re\lambda=-\frac{1}{2}\;,\qquad
  \Im\lambda\frac{2\alpha\sqrt{\Gamma}}{3\Gamma^{1/3}} \sign\varepsilon>0\; ,
\]
which lead to the correct choices of $\lambda$:
\begin{equation}
  \lambda =
    \cases{1 & : $\Phi_->0$\\
     -\frac{1}{2}+\rmi\frac{\sqrt{3}}{2}\sign\varepsilon & : $\Phi_-<0$}\;.
\end{equation}
Using this result, we can determine $\eta$ and $\delta$ from
(\ref{etaVonDelta}) and (\ref{deltaLsg}).
Equations (\ref{etaDeltaDef}) and (\ref{NFWirk}) then finally yield the 
desired parameter values
\begin{eqnarray}\label{abc_param}
\eqalign{ 
  c=-\left(\frac{\varepsilon}{6\eta}\right)^3 \;,\\
  a=3c^{2/3}\delta - \frac{\varepsilon}{8\Phi_{-1}}\;, \qquad
  b=3c^{2/3}\delta + \frac{\varepsilon}{8\Phi_{-1}} \; .
}\end{eqnarray}
Note that from \eref{abc_param} the parameters $a$, $b$, and $c$ are
explicitly determined as functions of the energy $\varepsilon$ and the
three action differences $\Phi_+$, $\Phi_-$, and $\Phi_{-1}$.

In our case, we will determine the normal form coefficients from the
scaled action differences shown in figure \ref{orbDat}.  To obtain the
actual non-scaled coefficients for different values of the magnetic
field strength $\gamma$, we need to derive scaling laws for the
coefficients. As we shall display the semiclassical spectra as
functions of scaled energy, we prefer not to scale the energy
difference $\varepsilon=\tilde{E}-\tilde{E}_c$. Then, with the help of
equations (\ref{NFWirk}) and (\ref{NFHess}), we can convince ourselves
that the scaling prescriptions
\begin{equation}
  \tilde{a}=\gamma^{-1/3}a\; , \qquad
  \tilde{b}=\gamma^{-1/3}b\; , \qquad
  \tilde{c}=\gamma^{-2/3}c
\end{equation}
fulfil the requirements of scaling actions according to
$\tilde{S}=\gamma^{1/3}S$ while not scaling Hessian determinants.

\section{Evaluation of the uniform approximation}
\label{EvalSec}
Now that the ansatz function $\Phi$ has been completely specified, a
suitable approximation to the coefficient $X$ in (\ref{unifApprox})
remains to be found. We assume $X$ to be independent of $\varphi$,
and as the value of $X$ is known at the stationary points of $f$ at
four different values of $I$ (including $I=0$), we approximate $X$ by
the third order polynomial $p(I)$ interpolating between the four given
points. This choice ensures that our approximation reproduces
Gutzwiller's isolated-orbits formula if, sufficiently far away from
the bifurcations, we evaluate the integral in
stationary-phase-approximation.
Thus, the uniform approximation takes its final form
\begin{equation}\label{UnifSol}
\fl d(E) = \frac{1}{2\pi^2m\hbar^2}\,\Re
     \exp\left\{\frac{\rmi}{\hbar}S_0(E)
                          -\rmi\frac{\pi}{2}\hat{\nu}\right\}
     \int dY\,dP_Y'\,p(I)
           \exp\left\{\frac{\rmi}{\hbar}\Phi(Y,P_Y')\right\}\;,
\end{equation}
which contains known functions only and can be evaluated numerically.

It is important to point out, at this juncture, the progress
encorporated in equation (\ref{UnifSol}) as compared to previous
literature results.  Starting from the lowest-order normal form
(\ref{NF2}), Sieber and Schomerus \cite{Sie98} derived a similar
formula for the contribution of an isolated generic
period-quadrupling bifurcation to the density of states.  However, as
we have seen in section \ref{generic_bif}, their normal form
(\ref{NF2}) describes the central periodic orbit and the stable and
unstable satellites only, from which it is evident that the uniform
approximation given by Sieber and Schomerus cannot take the presence
of the additional ghost orbit and the occurence of the ghost orbit
bifurcation into account.  The effect of the latter is negligible far
above the bifurcation energy $\tilde E_c$, where the asymptotic
behaviour is determined by the three real orbits common to both
(Sieber and Schomerus's and our own) forms of the uniform
approximation. The Sieber and Schomerus result, however, is not
capable of describing the correct asymptotic behaviour below $\tilde
E_c$, because one of the satellite orbits runs into a bifurcation
unforeseen by the normal form (\ref{NF2}) at $\tilde E_c'$, thus causing
Sieber and Schomerus's uniform approximation to diverge. Below $\tilde
E_c'$, when the stable satellite orbit does not exist any more, their
solution also ceases to exist because the required input data is no
longer available. Thus, to obtain a smooth interpolation between the
asymptotic Gutzwiller behaviour on either side of the bifurcations, we
must make use of the extended uniform approximation (\ref{UnifSol}),
which takes the contribution of the ghost orbit bifurcation into
account, as long as we do not go to the extreme semiclassical domain
where Planck's constant has become so small that below $\tilde E_c$
the asymptotic regime is reached before the ghost orbit bifurcation
can produce a palpable effect. It is only in this limit that the
impact of the ghost orbit bifurcation on the semiclassical spectrum
vanishes.

\begin{figure}
\vspace{20.0cm}
\includegraphics{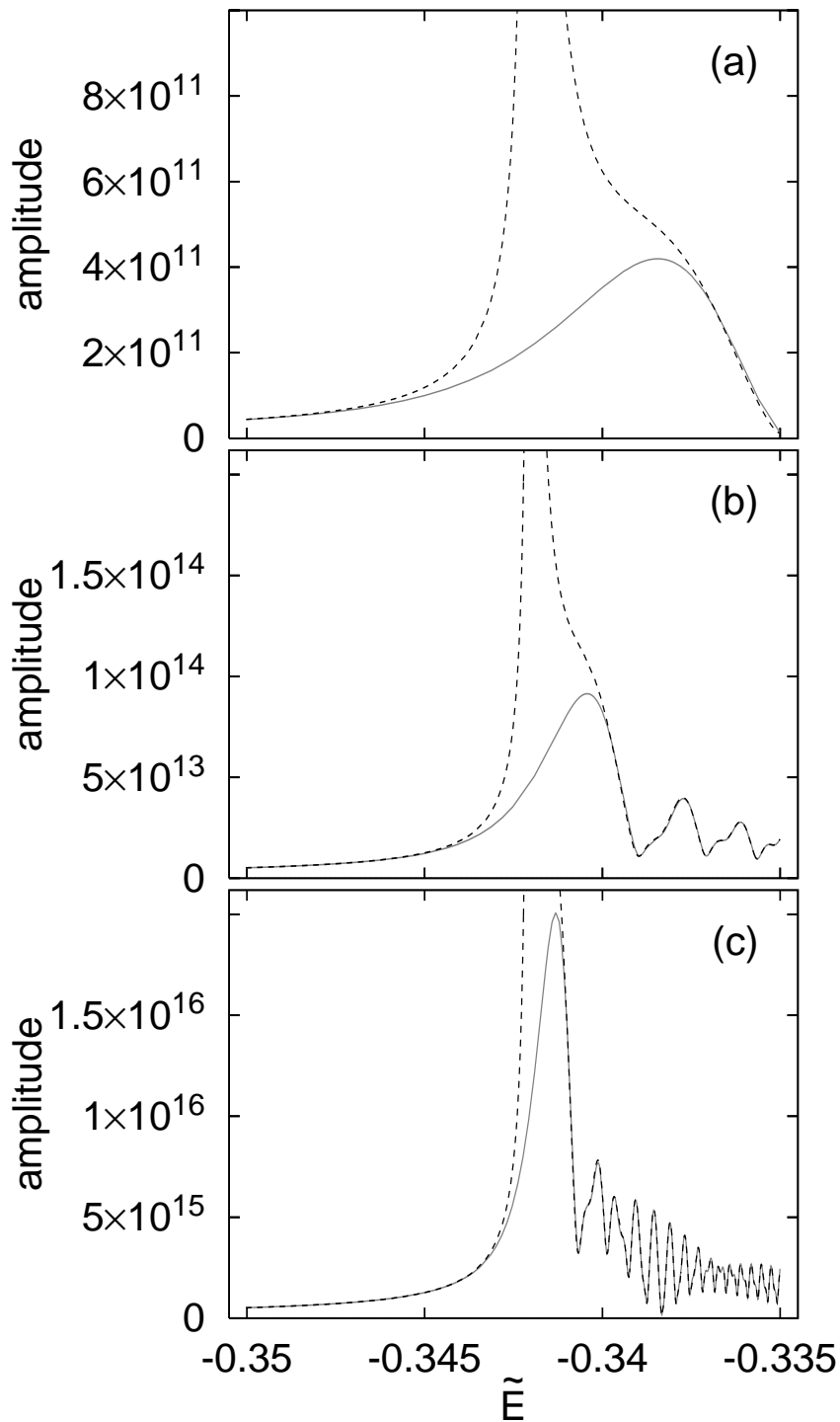}
\caption{\label{UnifFig}
Uniform approximation to the contribution of the considered bifurcations 
to the density of states for three different values of the magnetic field 
strength: (a) $\gamma=10^{-10}$, (b) $\gamma=10^{-12}$, (c) $\gamma=10^{-14}$.
Solid lines: Uniform approximations.
Dashed lines: Gutzwiller's trace formula.}
\end{figure}

We calculated the uniform approximation (\ref{UnifSol}) for three
different values of the magnetic field strength $\gamma$. The results
are shown in figure \ref{UnifFig}. To suppress the highly oscillatory
contributions originating from the factor
$\exp\left\{\frac{\rmi}{\hbar}S_0(E)\right\}$, we plot the absolute
value of (\ref{UnifSol}) instead of the real part.  As can be seen,
the uniform approximation proposed is finite at the bifurcation
energies, and, as the distance from the bifurcations increases,
asymptotically goes over into the results of Gutzwiller's trace
formula. Even the complicated oscillatory structures in the density of
states caused by interferences between the contributions from the
different real orbits involved at $\tilde{E}>\tilde{E}_c=-0.342025$
are perfectly reproduced by our uniform approximation (see figures
\ref{UnifFig}b and \ref{UnifFig}c). We also see that the higher the
magnetic field strength, the farther away from the bifurcation the
asymptotic (Gutzwiller) behaviour is acquired. In fact, for the
largest field strength in figure \ref{UnifFig}a ($\gamma = 10^{-10}$)
the asymptotic region is not reached at all in the energy domain
shown. The magnetic field dependence of the transition into the
asymptotic regime can be traced back to the fact that, due to the
scaling properties of our system, the scaling parameter $\gamma^{1/3}$
plays the r\^ole of an effective Planck's constant, therefore the
lower $\gamma$ becomes, the more accurate the semiclassical
approximation will be.

\section{Conclusion}

We have shown that in Hamiltonian systems with mixed regular-chaotic
dynamics {\em ghost orbit bifurcations} can occur
besides the bifurcations of real orbits. These are of special
importance when they
appear in the vicinity of bifurcations of real orbits, since 
they turn out to produce signatures in the
semiclassical spectra much the same as 
those of the real orbits. Consequently, the traditional theory
of uniform approximations for  bifurcations of real orbits
must be extended to also include the effects  of bifurcating ghost orbits.

We have illustrated the phenomenon of bifurcating ghost orbits in the 
neighbourhood of bifurcations of real orbits by way of example for 
the period-quadrupling of the balloon orbit in the diamagnetic Kepler
problem,  and have demonstrated how normal form theory can be extended
for this case so as to allow the unified description of both real 
{\em and} complex bifurcations.

We picked the example mainly for its simplicity, since (a) the real orbit
considered is one of the shortest fundamental periodic orbits in the 
diamagnetic Kepler problem and (b) the period-quadrupling is the 
lowest period-$m$-tupling possible ($m=4$) that exhibits the 
island-chain-bifurcation typical of all higher $m$. Thus we 
expect ghost orbit bifurcations to appear also for 
longer-period orbits, and, in particular, in the vicinity of 
all higher period-$m$-tupling bifurcations of real orbits. 

In fact, a general discussion of the bifurcation scenarios described
by the normal form (\ref{NF}) and its more general variant given in
\cite{Sad96} for different values of the parameters leads us to the
conclusion that the appearance of ghost bifurcations in the vicinity
of bifurcating real orbits is the rule, rather than the exception, in
general systems with mixed regular-chaotic systems, and thus one of
their generic features. It will be interesting and rewarding to study
higher period-$m$-tuplings with respect to the appearance of ghost
orbit bifurcations, and to extend ordinary normal form theory to also
include the contributions of ghost orbit bifurcations for all higher
$m$.

\section*{Acknowledgments}
This work was supported in part by the Sonderforschungsbereich No.~237
of the Deutsche Forschungsgemeinschaft. J.M.~is grateful to Deutsche
Forschungsgemeinschaft for a Habilitandenstipendium (Grant No.~Ma
1639/3).

\newpage

\end{document}